\documentstyle[prl,aps,twoside]{revtex}
\newcommand{\be}{\begin{equation}}
\newcommand{\ee}{\end{equation}}
\newcommand{\bea}{\begin{eqnarray}}
\newcommand{\eea}{\end{eqnarray}}
			       
\begin{document}

\draft
\title{Discrete approach to incoherent excitations in conductors}

\author{
  P. \v{Z}upanovi\'{c}$\mbox{ }^1 $, 
A. Bjeli\v{s}$\mbox{ }^2  $, 
\v{Z}. Agi\'{c}$\mbox{ }^1 $} 
 \address{
$\mbox{ }^1$  
 Department of Physics, Faculty of Science and Art,	       
University of Split\\
Teslina 10, 21000 Split, Croatia \\
$\mbox{ }^2 $ Department of Physics, Faculty of Science, University        
 of Zagreb\\
pp 331, 10002 Zagreb, Croatia}

\maketitle

\vspace{5mm}


\vspace{5mm}

\begin{center}
{\bf  Abstract}
\end{center}

Keeping the discreteness of the reciprocal space we calculate the spectrum of 
incoherent electron-hole excitations in the conducting Fermi liquids. The 
method is illustrated on the well-known jellium model within the random phase 
approximation. It also leads to the formulation of a sum rule from which we get 
the details of the dispersion curve for the collective plasmon mode. The notion 
of time averaging in the discrete approach is briefly recalled. 

\section{Introduction}   

The electron-hole excitations in conductors invoke in the frequency dependence 
of dielectric function $\epsilon ({\bf k}, \omega)$ a dense alternation of 
poles and zeros at the scale of discreteness of the reciprocal 
space \cite{pinesel}. Usually
this complex mathematical problem is avoided by making the standard procedure 
of continuation of the wave vector variable {\bf k} which is well founded for 
macroscopic systems, and in addition includes the standard proposition allowing 
for the irreversibility in the thermodynamic limit \cite{doniach}. By this  
the dense set of poles and zeros is eliminated and $\epsilon ({\bf k}, \omega)$
becomes an analytic functions. In particular, within random phase approximation 
(RPA) for jellium which will be followed here, ${\em Im}\,\epsilon$ is a continuous 
function of ${\bf k}$ and $\omega$ in the range of variables covering the 
so-called electron-hole continuum.

Although cumbersome at first sight, the discrete presentation of dielectric 
response still appears to be advantageous and physically more transparent in 
some treatments, like in the calculations of cohesive energy \cite{zbb,zb} and 
one-particle spectral function \cite{zab} in the single band and multiband 
systems. In the present work we show how the method developed in
Refs.\cite{zb,zab} reproduces some well-known results for the jellium model.
In particular we derive an explicit expression  for energies of incoherent 
excitations, which is of course not attainable after continuation. This leads 
to the formulation of a sum rule which enables a direct determination of the 
dispersion curve for the collective excitation (i. e. plasmon in the present 
example). The particular detail which then can be followed in a transparent way 
is the cross-over of the plasmon dispersion from the collective to the 
incoherent regime as the wave vector increases, the subject which was often 
exposed incorrectly or imprecisely in literature. 

In Section II we start with the formulation of the problem and continue with 
the explicit calculation of the energies of incoherent excitations. Section III 
contains the formulation of the sum rule, the analysis of the plasmon 
dispersion and the short note on the calculation of ${\em Im} \,\epsilon$ in the 
discrete approach. Concluding remarks are given in Section IV.    

\section{Incoherent excitations}

We start from the well-known RPA dielectric function for the jellium 
\cite{pinesel}. The excitation energy for a given value of the wave vector  
{\bf q} is the solution of the equation $\varepsilon({\bf q},\omega)=0$ in the 
frequency variable $\omega$. Let us write this equation in the form
\be
\label{dielectric1}
\varepsilon({\bf q},\omega)=
1-\frac{4 e^2}{\pi q^2 L} \sum_{k_{\parallel}=-k_F}^{k_F}
g( k_{\parallel}, {\bf q})\frac{E(k_{\parallel},{\bf q})}
{[\omega+i \mbox{sign}(\omega)\eta]^2 - E^2(k_{\parallel},{\bf q})}=0,
\ee
where $L^3$ is the volume of the system. We keep throughout this text the 
discrete summation in terms of the wave vector component parallel to {\bf q}, 
\be
\label{k}
k_{\parallel}=n\frac{2\pi}{L} \;\;,\;\; n \in Z.
\ee
The equation $k_{\parallel}=\mbox{constant}$ defines the locus of the constant
electron-hole excitation energy
\be
\label{E}
E(k_{\parallel},{\bf q})=\frac{1}{2m}(2k_{\parallel}+q)q , 
\ee
i.e. the summation in Eq.(\ref{dielectric1}) goes over all such loci. As shown 
in Fig.\ref{loci} the locus for given values of $q$ and $k_{\parallel}$ is 
either circular [for $E(k_{\parallel},{\bf q}) > (k_F^2-k^2_{\parallel})/2m$]
or annular [for $E(k_{\parallel},{\bf q}) < (k_F^2-k^2_{\parallel})/2m$]
surface, centered at $k_{\parallel}$. The radius of the former is 
$k_{\perp M} \equiv \sqrt{k_F^2-k^2_{\parallel}}$, while the latter is bounded 
by concentric circles with radii $k_{\perp M}$ and 
$k_{\perp m} \equiv \sqrt{k_F^2-(k_{\parallel}+q)^2}$.

The number of the points multiplied with the elementary surface $(2 \pi/L)^2$ 
of the reciprocal space in a given locus is
\begin{equation}
\label{smallq}
g(k_{\parallel},{\bf q})= \left\{
\begin{array}{ll}
(k_F^2-k_{\parallel}^2)\pi &  k_F > k_{\parallel} > k_F -q \\
(2k_{\parallel}+q)q\pi &  k_F-q > k_{\parallel} > -q/2
\end{array} \right.
\mbox{for}\;\;  q < 2k_F, \;\;
\ee
  and
  \be
\label{bigq}
   g(k_{\parallel},{\bf q})=
  (k_F^2-k_{\parallel}^2)\pi
\;\; \mbox{   for} \;\;  q > 2k_F.
\end{equation}
                       
The dielectric function (\ref{dielectric1}) diverges at energies which are
infinitesimally close to the values of electron-hole excitation energies 
(\ref{E}) at the real $\omega$-axis. The energy difference between neighboring 
poles (\ref{E}) is 
\be
\label{delta}
\Delta E({\bf q})=E(k_{\parallel}+2\pi/L,{\bf q}) -E(k_{\parallel} ,{\bf q}) =
\frac{2\pi q}{mL}.
\ee
Between each such pair of neighboring poles (\ref{E}) there should be a zero 
of dielectric function. In other words all solutions of the equation 
$\varepsilon({\bf q},\omega)=0$ except the largest one lie between neighboring 
electron-hole energies. Let us denote these zeros by 
$\Omega(k_{\parallel},{\bf q})$, and write 
\be
\label{incoherent}
\Omega(k_{\parallel},{\bf q}) + i  \mbox{sign}(\Omega)\eta=
E(k_{\parallel},{\bf q}) +
\Theta(k_{\parallel},{\bf q})\Delta E({\bf q})
\ee
with $0<\Theta(k_{\parallel},{\bf q}) <1$. Let us remind that in contrast to 
these {\em incoherent excitations}, the excitation with largest energy can be 
macroscopically (with respect to $1/L$) far from its closest electron-hole 
counterpart, $E(k_F,{\bf q})$. Such isolated zero is a 
{\em collective excitation}, i. e. plasmon in the present jellium model.

An approximate expression for the energies of incoherent excitations follows 
after retaining the $\omega$-dependence only in those terms of 
Eq.(\ref{dielectric1}) which contain nearest neighboring poles to a given zero 
$\Omega(k_{\parallel},{\bf q})$, i. e. only in terms with 
$E(k_{\parallel},{\bf q})$ and $E(k_{\parallel}+2\pi/L,{\bf q})$. In terms 
with wave vectors different from $k_{\parallel}$ or $k_{\parallel}+2\pi/L$ we 
substitute $\omega + i \mbox{sign}(\omega) \eta$ with
$E(k_{\parallel},{\bf q})$. The validity of this approximative step is 
discussed in the Appendix. Eq.(\ref{dielectric1}) now reduces to 
\be
\label{dysapprox}
1-\frac{4}{\pi^2q^2 a_0}\left\{F(k_{\parallel},{\bf q})+
\frac{\pi}{mL} \left[\frac{g(k_{\parallel},
{\bf q})
E(k_{\parallel},{\bf q})}{[\omega +  i \mbox{sign}(\omega)\eta]^2 -
E^2(k_{\parallel},{\bf q})}
+  \frac{g(k_{\parallel}+\frac{2\pi}{L},{\bf q})
E(k_{\parallel}+\frac{2\pi}{L},{\bf q})}{[\omega + i \mbox{sign}(\omega)\eta]^2
-E^2(k_{\parallel}+\frac{2\pi}{L},{\bf q})}\right] \right\}=0,
\ee
where $a_0=1/me^2$ is Bohr radius and 
\be
\label{F}
F(k_{\parallel},{\bf q})=\frac{\pi}{mL}
\left[ \sum_{k'_{\parallel}=-k_F}^{k_{\parallel}-2\pi/L}
\frac{g(k'_{\parallel},{\bf q})
E(k'_{\parallel},{\bf q})}{E^2(k_{\parallel},{\bf q})
-E^2(k'_{\parallel},{\bf q})}
+  \sum^{k_F}_{k'_{\parallel}=k_{\parallel}+4\pi/L}
\frac{g(k'_{\parallel},{\bf q})
E(k'_{\parallel},{\bf q})}{E^2(k_{\parallel},{\bf q})
-E^2(k'_{\parallel},{\bf q})}\right]. 
\ee

With the most divergent terms singled out, we can now make continuation of the 
residual sums (\ref{F}), by replacing each term with the integral on the 
interval $2\pi/L$. Choosing to replace a term characterized by the discrete 
value $k'_{\parallel}$ with the integral from $k'_{\parallel}-2\pi/L$ to 
$ k'_{\parallel}$ we come to the expression 
\be
\label{fint}
F(k_{\parallel},{\bf q})=\frac{1}{2m}
\left[   
\int^{k_{\parallel}-2\pi/L }_{-k_F} d\;k'_{\parallel}
\frac{g(k'_{\parallel},{\bf q})
E(k'_{\parallel},{\bf q})}{E^2(k_{\parallel},{\bf q})
-E^2(k'_{\parallel},{\bf q})} +
\int^{k_F}_{k_{\parallel}+2\pi/L } d\;k'_{\parallel}
\frac{g(k'_{\parallel},{\bf q})
E(k'_{\parallel},{\bf q})}{E^2(k_{\parallel},{\bf q})
-E^2(k'_{\parallel},{\bf q})}\right],
\ee
where $k_{\parallel}-2\pi/L$ and $k_{\parallel}+2\pi/L $ are the lower and 
upper boundaries of the interval that corresponds to the missing  terms 
$k_{\parallel}$ and $k_{\parallel}+2\pi/L $ in the sum (\ref{F}). We note
that the final result of integration in Eq.(\ref{fint}) does not depend on the 
choice of integration boundaries for $2\pi/L$ intervals, as specified above. 
  
Since the primitive function in the above integral is odd with respect to the 
pole at $k_{\parallel}$, the expression (\ref{fint}) can be rewritten in the 
form 
\be
\label{Fint}
F(k_{\parallel},{\bf q})=\frac{1}{2m}
\int^{k_F}_{-k_F} d\;k'_{\parallel}\frac{g(k'_{\parallel},{\bf q})
E(k'_{\parallel},{\bf q})}{E^2(k_{\parallel},{\bf q})
-E^2(k'_{\parallel},{\bf q})}.
\ee
After elementary integration we get
\be
\label{f}
F(k_{\parallel},{\bf q}) =
\frac{\pi}{4} \left[-2k_F - \frac{k_F^2-k_{\parallel}^2}{q}
\ln\left|\frac{(k_F-k_{\parallel})(k_F+k_{\parallel}+q)}
{(k_F+k_{\parallel})(k_F-k_{\parallel}-q)}\right|+(2k_{\parallel}+q)
\ln \left| \frac{k_F+k_{\parallel}+q}{ k_F-k_{\parallel}-q} \right| \right].
\ee
Furthermore, by solving the biquadratic equation (\ref{dysapprox}) we get for 
the function $\Theta(k_\parallel, {\bf q})$ in Eq. (\ref{incoherent}) the 
expression
\be
\label{difom}
\Theta(k_\parallel, {\bf q}) = \frac{1}{2}[1+x -\mbox{sign}(x)\sqrt{1+x^2}],
\ee
where 
\be 
\label{x}
 x \equiv \frac{2 g(k_{\parallel},{\bf q})}
{q[  \pi^2q^2 a_0-4 F(k_{\parallel},{\bf q})]}.
\ee
Evidently $\Theta$ varies in the range $\left(0,1\right)$ in accordance with  
the starting assumption. More precisely 
 \be
\label{omregion}
\begin{array}{lll}
0 <    \Theta(k_{\parallel},{\bf q})
 < 
1/2 \\
    \Theta(k_{\parallel},{\bf q}) =1 \\
1/2  <    \Theta(k_{\parallel},{\bf q})
 <1
\end{array}
\mbox{for }  \pi^2q^2 a_0 - 4 F(k_{\parallel},{\bf q})
\begin{array}{l}
>0\\
=0 \\
<0
\end{array} .
\ee

The above results can be written in a more explicit way in the limits 
$q << k_F$ and $ q >> 2 k_F$. In the former case and for 
$ k_{\parallel} \ll k_F$ the expression (\ref{f}) reduces to
\be
\label{small}
F(k_{\parallel},{\bf q})  \approx
-\pi k_F + \frac{\pi k_{\parallel}^{2}}{2 k_F}+
\frac{\pi(2k_{\parallel}+q)(k_{\parallel} +q)}{2 k_F} +
\ee
and the leading term in the energy of incoherent excitations is  
\be
\Omega(k_{\parallel},{\bf q})=E(k_{\parallel},{\bf q})+
\frac{\pi q(2k_{\parallel}+q)}{2 k_FmL} ,
\ee
i.e. the zeros of $\epsilon(\omega, {\bf k})$ in the $\omega$-plane lie very
close to the corresponding electron-hole excitation poles. Here and further on
we omit for simplicity the infinitesimal imaginary part in excitation energies
(\ref{incoherent}).

As the wave number $k_{\parallel}$ increases the energy of incoherent 
excitations moves gradually towards the first higher neighboring 
electron-hole excitation. For $k_{\parallel}$ close to $k_F$ and for 
$k_F \gg q \gg k_F -k_{\parallel}$ the function (\ref{f}) reduces to 
\be
\label{kf}
F(k_{\parallel},{\bf q}) \approx  \frac
{\pi(2 k_F+q) }{4}
\ln \frac{2k_F+q}{q}-\frac{\pi k_F}{2}-
\frac{\pi k_F(k_F-k_{\parallel})}{2q} \ln \frac{k_F- k_{\parallel}}{q} .  
\ee
Keeping only the leading term in the above expression the approximate 
expression for corresponding energies of incoherent excitation reads 
\be
\label{larger}
\Omega(k_{\parallel},{\bf q})= E(k_{\parallel} + 2\pi/L,{\bf q})- 
\frac{2\pi(k_F-k_{\parallel})}{mL\ln (2k_F/q)} ,
\ee
i. e. these energies are very close to the energies of the first larger 
neighboring electron-hole excitations.

In the latter case $q >> 2 k_F$ Eq.(\ref{f}) reduces to
\be
\label{large}
4 F(k_{\parallel},{\bf q}) \approx \pi\left(\frac{2k_{\parallel} k_F}{q}
- \frac{k_F^2- k^2_{\parallel}}{q}\ln
\frac{  k_F- k_{\parallel}}{k_F+ k_{\parallel}} \right) \ll  \pi^2q^2 a_0.
\ee
Energies of incoherent excitations are then given by 
\be
\label{largeq}
\Omega^2(k_{\parallel},{\bf q})-E^2(k_{\parallel},{\bf q})=
\frac{4 e^2}{\pi L q^2}g(k_{\parallel},{\bf q})
\left[1+ \frac{4me^2 F(k_{\parallel},{\bf q})}{\pi^2 q^2}\right]
 E(k_{\parallel},{\bf q}),
\ee
i.e. they are very close to their electron-hole counterparts for all values 
of $k_{\parallel}$. The physical reason for this small departure of 
incoherent excitations from bare electron-hole ones is the weakness of the 
bare Coulomb interaction $4\pi e^2/q^2$ in this regime.

It follows from the foregoing analysis that the regime of small values of $q$ 
qualitatively differs from that of large ones. 
In the former case the energy of incoherent excitations moves from their 
electron-hole counterpart towards the first larger neighboring 
electron-hole excitation, crossing the half-width $\Delta E({\bf q})/2$  
(Eq. \ref{delta}), as $k_{\parallel}$ moves from the lower bound, $-q$, to 
the upper bound, $k_F -2\pi/L$. On contrary, in the latter case it does not 
cross the half-width. It is interesting to estimate the value of the wave 
vector $q_{cr}$ which roughly separates these two regimes. To this end we 
note that in the regime of  small values of $q$ the difference between the 
energy of incoherent excitation and its electron-hole counterpart has the 
largest value for $k_{\parallel} \approx k_F$. Let us define the critical 
wave number $q_{cr}$ for which $\Theta(k_F,q_{cr})$ (\ref{difom}) is equal to 
$1/2$. In this case $x$ in Eq. (\ref{x}) diverges, i.e. $q_{cr}$ is the 
solution of the equation 
\be
\label{krit}
\pi q_{cr}^2 a_0=
 -2k_F+(2k_F+q_{cr})\ln \frac{2k_F+q_{cr}}{q_{cr}} .
\ee

To summarize, for $q>q_{cr}$ the energies of incoherent excitations are closer 
to their electron-hole counterparts than to the next neighboring electron-hole 
excitations, regardless to the value of the wave number $k_{\parallel}$. For 
$q<q_{cr}$ the energies of incoherent excitations are closer either to their 
electron-hole counterparts or to the first larger neighboring electron-hole 
excitations for small and large values of the wave number $k_{\parallel}$
respectively. The wave number $q_{cr}$ roughly separates these two regimes. 

\section{Collective excitations}

In the above approximate calculations of energies of incoherent excitations 
we explore the fact that corresponding zeros and poles of the dielectric 
function densely alternate on the $\omega$-axis. This method however cannot be 
used for the excitation with the highest energy, lying above the maximum of 
electron-hole excitation energy $E(k_F, {\bf q})$ for given ${\bf q}$. 

In order to determine the $\Omega(k_F, {\bf q})$-dependence of this collective 
mode, we write Eq. (\ref{dielectric1}) in the form 
\be
\label{dyseq3}
\frac{\sum_{k_{\parallel}=-k_F}^{k_F}g(k_{\parallel},{\bf q})E(k_{\parallel},
{\bf q})
\prod_{k'_{\parallel} \neq k_{\parallel}}\left[\omega^2 -
E^2(k'_{\parallel},{\bf q}) \right]-
(q^2 L\pi /4 e^2)\prod_{k_{\parallel}=-k_F}^{k_F}
\left[\omega^2 -E^2(k_{\parallel},{\bf q})\right]}
{\prod_{k_{\parallel}=-k_F}^{k_F}
\left[\omega^2 -E^2(k_{\parallel},{\bf q})\right]} =0.
\ee
Factors that multiply the highest powers, $\omega^{k_FL\pi}$ and
$\omega^{k_FL\pi-2}$ in the polynomial presentation of the nominator on the
left-hand side are
\be
\label{an}
a_{k_F} =  -q^2 L\pi /4 e^2,
\ee
and 
\be
\label{an-1}
a_{k_F-2\pi/L} =  \sum_{k_{\parallel}=-k_F}^{k_F}
g(k_{\parallel},{\bf q})E(k_{\parallel},{\bf q}) + \frac{q^2 L \pi }{4 e^2}
\sum_{k_{\parallel}=-k_F}^{k_F} E^2(k_{\parallel},{\bf q})
\ee
respectively. As it follows from the elementary algebra the ratio 
$a_{k_F-2\pi/L}/a_{k_F}$ is equal to the sum of all zeroes of the 
Eq. (\ref{dyseq3}). From this we get
\be
\label{viete}
\sum_{k=-k_F}^{k_F-2\pi/L}\Omega^2(k_{\parallel},{\bf q})
- E^2(k_{\parallel},{\bf q}) + \Omega^2(k_F,{\bf q})-E^2(k_F,{\bf q})=
\omega_{pl}^{2}.
\ee

Above relation is a sum rule which states that the sum of differences of the 
squares of excitation energies and corresponding electron-hole energies 
equals to the square of the plasmon energy 
$\omega_{pl}^{2}=4\pi Ne^2/(L^3m)$ \cite{pinesel}. The dispersion of the
collective mode $\Omega(k_F,{\bf q})\, \equiv \Omega_{pl}({\bf q})$  
follows directly from this relation once the energies of the incoherent 
excitations are determined, as it was done in the previous Section. To 
illustrate, let us recall two characteristic points of the dispersion curve 
$\Omega_{pl}({\bf q})$. First, since all incoherent and electron-hole 
excitations vanish for ${\bf q}=0$, the above sum rule reproduces the 
well-known result that $\omega_{pl}$ is the energy of the highest 
long-wavelength excitation, i. e. of the plasmon \cite{pinesel}. Second, it 
is known from the continuum approach \cite{works} that the dispersion curve 
$\Omega_{pl}({\bf q})$ touches the border of the electron-hole continuum at a 
finite wave number. Here we point out that this wave number {\em coincides} 
with the wave number $q_{cr}$ given by Eq. (\ref{krit}). The result by which 
plasmons do not exist as collective excitations  at  wave numbers at 
which all incoherent excitations are closer to their electron-hole 
counterparts than to next neighboring electron-hole excitations can be 
derived in the following way.
The plasmon dispersion $\Omega_{pl}({\bf q})$ follows also from 
Eq. (\ref{dysapprox}) after omitting the term with poles and substituting  
$E(k_{\parallel},{\bf q})$ by $\Omega_{pl}({\bf q})$ in the expression 
(\ref{F}) for the function $F(k_{\parallel}, {\bf q})$. Thus we get 
\be
\label{plasmonq}
1-\frac{4}{\pi^2q^2 a_0}
F\left(\frac{m\Omega_{pl}({\bf q})}{q}-\frac{q}{2},{\bf q}\right)=0.
\ee
In the continuum approximation the plasmon dispersion touches the border of 
the electron-hole quasi-continuum for $\Omega_{pl}({\bf q})= (2k_F+q)q/2m$.
By this Eq. (\ref{plasmonq}) reduces to Eq. (\ref{krit}), i. e. the plasmon 
indeed ceases to exist as a collective excitation just at $q=q_{cr}$.
        
In the original works \cite{works} as well as in the 
textbooks \cite{pinesel,doniach,pinnoz,fetter,mahan} the touching point 
$q_{cr}$ is usually interpreted as the wave number above which the decay of 
plasmons into electron-hole pairs takes part. Sometimes it is even stated or 
hinted that the dispersion curve $\Omega_{pl}({\bf q})$ and the upper border of 
electron-hole range cross at $q=q_{cr}$  \cite{doniach,mahan}. We note that 
within the present discrete approach the curve $\Omega_{pl}({\bf q})$ neither 
touches nor crosses the border of electron-hole excitations, but just 
approaches it at the distances of the order of $\Delta E({\bf q})$, and 
remains at this distance for $q>q_{cr}$. To show this quantitatively let us 
calculate the highest excitation energy for $q \gg q_{cr}$ by using the 
expression (\ref{largeq}) for the differences 
$\Omega^2(k_{\parallel}, {\bf q}) - E^2(k_{\parallel}, {\bf q})$ 
in this limit. We get from the sum rule (\ref{viete}) 
\be  
\label{incpl1}
\Omega^2(k_F,q)-E^2(k_F,q)=\omega_{pl}^2-
\frac{2e^2}{Lmq}\sum_{k_{\parallel} \neq k_F}
 \left(k_F^2-k_{\parallel}^2\right)\left(2k_{\parallel}+q\right)
\left[1+ \frac{4me^2 F(k_{\parallel},{\bf q})}{\pi^2 q^2}\right].
\ee
In order to calculate the sum on the right-hand side we devide it in the convenient way and make the continuation,
\be
\label{incpl2}
\Omega^2(k_F,q)-E^2(k_F,q)=\omega_{pl}^2-\frac{e^2}{ \pi mq}\left[ \int_{-k_F}^{k_F}dk
 \left(k_F^2-k^2\right)\left(2k+q\right)
\left[1+ \frac{4me^2 F(k_{\parallel},{\bf q})}{\pi^2 q^2}\right]
- \int_{k_F-2\pi/L}^{k_F}dk
 \left(k_F^2-k^2\right)\left(2k+q\right) \right].
\ee
After straightforward steps we get
\be
\label{incpl3}
\Omega(k_F,q)=E(k_F,q)+\frac{e^2k_F}{\pi q^2}\left(\frac{2\pi}{L}\right)^2,
\ee
noting that corrections to the energies of incoherent excitations as calculated in the previous Section 
[introduced in Eq.(\ref{incpl2}) through the function 
$ F(k_{\parallel},{\bf q})$] do not contribute to the result (\ref{incpl3}) 
up to the order of $\omega^4_{pl}/(q^6/m^3)$.

To conclude, the above analysis shows that the highest branch 
of excitations gradually approaches the quasi continuous incoherent
electron-hole range as $q$ approaches $q_{cr}$ from below. In this range it
is a well-defined collective (plasmon) mode. For $q > q_{cr}$ it
remains above the top of this range at the microscopic energy difference of 
the order of
$\Delta E({\bf q})$ or less, and as such does not have the properties of
a collective mode. Simultaneously as $q$ passes through $q_{cr}$
qualitative changes in the incoherent electron-hole range take place, as was
already emphasized at the end of Sec. II. In Fig.\ref{plazmon} we illustrate 
the above discussion, which to some extent complements that from 
Ref.\cite{pinnoz}, with numerical results for a large finite system. Note 
that our plasmon dispersion clearly differs from those schematically presented 
in e. g. Refs.\cite{doniach,mahan}.
  
We close this Section with a short remark, based on the arguments given in 
Ref. \cite{doniach}, on the calculation of 
${\em Im} \varepsilon({\bf q},\omega)$ in the present discrete approach. In 
order to get a proper result for the dissipative contributions to the 
correlation and response functions, one uses the standard recipe, i. e. 
calculates imaginary parts of their Fourier transforms only after the 
continuation in the reciprocal {\bf q} - space. This order of steps is based 
on the assumption that the characteristic energetic level spacing in the 
system is sufficiently small in comparison  to the reciprocal time of 
observation of the system. The equivalent proposition in the classical 
statistical physics is that the available time for the statistical averaging 
is much shorter than the Poincare cycle time. The duration of the time of 
observation however becomes irrelevant (i. e. it can be assumed arbitrarily 
long), once the continuation is performed. 

If one keeps, like in the present approach, the discrete summations
throughout the calculations, the dissipative contributions are well-defined 
only after making averaging on the energy scales larger than the inherent 
energy level spacings. In particular, in our case one reproduces the correct 
result for ${\em Im} \varepsilon({\bf q},\omega)$ by averaging in $\omega$ on an 
interval not smaller than the energy differences $\Delta E({\bf q})$ from 
Eq. (\ref{delta}). Indeed after averaging the imaginary part of the dielectric  
function (\ref{dielectric1}),
\be
\label{imagdis}
{\em Im} \varepsilon({\bf q},\omega)=
\frac{2 e^2}{q^2 L} \sum_{k_{\parallel}=-k_F}^{k_F}
g( k_{\parallel}, {\bf q})
\left\{\delta\left[
\omega+ E(k_{\parallel},{\bf q})\right]+\delta\left[
\omega- E(k_{\parallel},{\bf q})\right]\right\},
\ee
over $\Delta E({\bf q})$, the smallest possible energy interval consistent 
with the above proposition, we get
\be
\label{imagcon}
\overline{ {\em Im} \varepsilon({\bf q},\omega)}=
\frac{1}{\Delta E({\bf q})} 
\int_{\omega-\Delta E({\bf q})/2}^{\omega +\Delta E({\bf q})/2 }
{\em Im} \varepsilon({\bf q},\omega') d \omega'
=\frac{m e^2}{\pi q^3}
g(\frac{m \omega}{q}-\frac{q}{2} , {\bf q}) \,.
\ee
This is the well-known result for the imaginary part of the Lindhard 
function \cite{pinesel,fetter,mahan} in the region of the electron-hole quasi 
continuum. We note that the above averaging procedure might be particularly 
relevant for mesoscopic situations in which the cross-over from the dissipative 
regime to the regime without dissipation by varying the width of time (i. e. 
frequency) window becomes attainable experimentally.

\section{Conclusion}

In the standard treatments with continuous wave vector the electron-hole 
region is reduced to a structureless continuum. The present analysis in which 
we keep systematically the wave vector discreteness leads to explicit results 
for shifts of energies of incoherent excitations with respect to the 
corresponding bare electron-hole excitation energies. Although these shifts 
are infinitesimal in macroscopic systems, their knowledge enables non-standard
calculations of other physically observables like correlation 
energies \cite {zb} and spectral functions \cite{zab}. The simple example of 
this kind is the calculation of the plasmon dispersion through the use of the 
sum rule (\ref{viete}). 

We note that the approach presented here may be straightforwardly extended 
to more complex macroscopic (e. g. multiband\cite {zb,zab}) systems. 
Furthermore, it is obviously particularly appropriate in studies of small 
mesoscopic systems in which electron-hole excitations are characterized by 
essentially larger energy level spacings with respect to those in the 
macroscopic limit. Then one also encounters interesting possibility of 
cross-over from the dissipatively irreversible to reversible regime, connected 
with fundamental principles of thermodynamic averaging.

\bigskip

\appendix{ \begin{center}
 APPENDIX  \end{center}}

\bigskip

In this Appendix we consider the validity of the approximation introduced 
by passing from Eq. ({\ref{dielectric1}) to Eq. (\ref{dysapprox}). By this step we 
replace in the residual sums [represented by the function 
$F(k_{\parallel}, {\bf q})$ (Eq. \ref{f})] the exact value of a given solution 
$\Omega(k_{\parallel}, {\bf q})$ of Eq. ({\ref{dielectric1}) by the corresponding 
pole $E(k_{\parallel}, {\bf q})$ as defined by Eq. (\ref{incoherent}). Let us 
start from the exact expression for the function $F(k_{\parallel}, {\bf q})$ 
\small
\be
\label{Fexa}
\tilde{F}(k_{\parallel},{\bf q})=
\frac{\pi}{mL} \left[ \sum_{k'_{\parallel}=-k_F}^{k_{\parallel}-2\pi/L}
\frac{g(k'_{\parallel},{\bf q})
E(k'_{\parallel},{\bf q})}{\left[E(k_{\parallel},{\bf q})+
\Theta(k_{\parallel},{\bf q})\Delta E({\bf q} )\right]^2
-E^2(k'_{\parallel},{\bf q})}
+  \sum^{k_F}_{k'_{\parallel}=k_{\parallel}+4\pi/L}
\frac{g(k'_{\parallel},{\bf q})
E(k'_{\parallel},{\bf q})}{ \left[E(k_{\parallel},{\bf q})+
\Theta(k_{\parallel},{\bf q} )\Delta E({\bf q}) \right]^2
-E^2(k'_{\parallel},{\bf q})} \right].
\ee
\normalsize
Here the value of the zero of Eq.(\ref{dielectric1}) is written in the form 
(\ref{incoherent}). The Taylor expansion of $\tilde{F}(k_{\parallel},{\bf q})$
in terms of $\Theta(k_{\parallel},{\bf q})\Delta E({\bf q})$ gives 
\be 
\label{Taylor}
\tilde{F}(k_{\parallel},{\bf q})-F(k_{\parallel},{\bf q})=
F_1(k_{\parallel},{\bf q})\Theta(k_{\parallel},{\bf q})\Delta E({\bf q}) +
 {\cal O}\left[[\Theta(k_{\parallel},{\bf q})\Delta E({\bf q})]^2\right], 
\ee
with 
\be
\label{kor1}
\label{F1}
F_1(k_{\parallel},{\bf q})=
-\frac{2\pi}{mL}
E(k_{\parallel},{\bf q})
\left[ \sum_{k'_{\parallel}=-k_F}^{k_{\parallel}-2\pi/L}
\frac{g(k'_{\parallel},{\bf q})
E(k'_{\parallel},{\bf q})}{\left[E^2(k_{\parallel},{\bf q})
-E^2(k'_{\parallel},{\bf q})\right]^2}
+  \sum^{k_F}_{k'_{\parallel}=k_{\parallel}+4\pi/L}
\frac{g(k'_{\parallel},{\bf q})
E(k'_{\parallel},{\bf q})}{\left[E^2(k_{\parallel},{\bf q})^2
-E^2(k'_{\parallel},{\bf q})\right]^2}\right].
\ee
This correction can be estimated after replacing sums by integrals. We get
\be
F_1(k_{\parallel},{\bf q}) = -\frac{g(k_{\parallel},{\bf q})}
{2q\Delta E({\bf q})}.
\ee

The approximation is justified if 
\be
\label{crit}
\left| \frac{F_1(k_{\parallel},{\bf q})
\Delta E({\bf q})}{F(k_{\parallel},{\bf q})}\right| =
\frac{2g(k_{\parallel},{\bf q})}{\pi\left|
-2k_Fq - (k_F^2-k_{\parallel}^2)
\ln\left|\frac{(k_F-k_{\parallel})(k_F+k_{\parallel}+q)}
{(k_F+k_{\parallel})(k_F-k_{\parallel}-q)}\right|+(2k_{\parallel}+q)q
\ln \left| \frac{k_F+k_{\parallel}+q}{ k_F-k_{\parallel}-q} \right|
\right|}\ll 1.
\ee
This condition is best fulfilled for small and large excitation energies due to 
the vanishing density of electron-hole excitations $g(k_{\parallel},{\bf q})$. 
Eq. (\ref{crit}) also indicates that the approximation might be less adequate in 
the range of wave numbers k in which $F(k,{\bf q})=0$ is small. The value $k_0$
for which $F(k_0,{\bf q})=0$ can be determined from Eq. (\ref{f}). Assuming  
$q \ll k_F-{k_0}$ we get 
\be
\frac{k_0}{k_F} \ln \frac{1+\frac{k_0}{k_F}}{1-\frac{k_0}{k_F}} \approx 1,
\ee
and $k_0\approx 0.65 k_F$. 

The direct insight into the validity of method follows from the comparison of 
analytically calculated values of differences $\Theta( k_{\parallel}, {\bf q})=
[\Omega( k_{\parallel}, {\bf q})-E(k_{\parallel}{\bf q})]/\Delta E({\bf q})$ 
and those obtained numerically for the mesoscopic system of 
$N \approx 8 \cdot 10^6\pi/3$ electrons, and  $L/a_0=10$, which by means of 
Eq. (\ref{krit}) gives  $ q _{cr} \approx 0.13 k_F$. It is shown in 
Fig.\ref{test}. For $q = 0.1 k_F < q_{cr}$ this difference raises monotonously 
from zero to unity, while for $q = 0.3 k_F > q_{cr}$ it shows highly 
non monotonous behavior. Finally for $q=2k_F >> q_{cr}$ $\Theta( k_{\parallel}, 
{\bf q})$ is close to zero in the whole range of wave numbers $k_{\parallel}$. 
The curves in Fig. \ref{test} clearly confirm the above estimations that the 
deviations of the results of analytic method from Sec. II from numerical 
calculations are negligible for small and large excitation energies. For 
intermediate values of excitation energies and for $q < q_{cr}$ the method is 
less accurate, but deviations from numerical results are still quite small. 
Finally for $q >> q_{cr}$ this deviation is negligible for all excitations.

\newpage

\begin{figure}
\vspace*{18cm}
\includegraphics{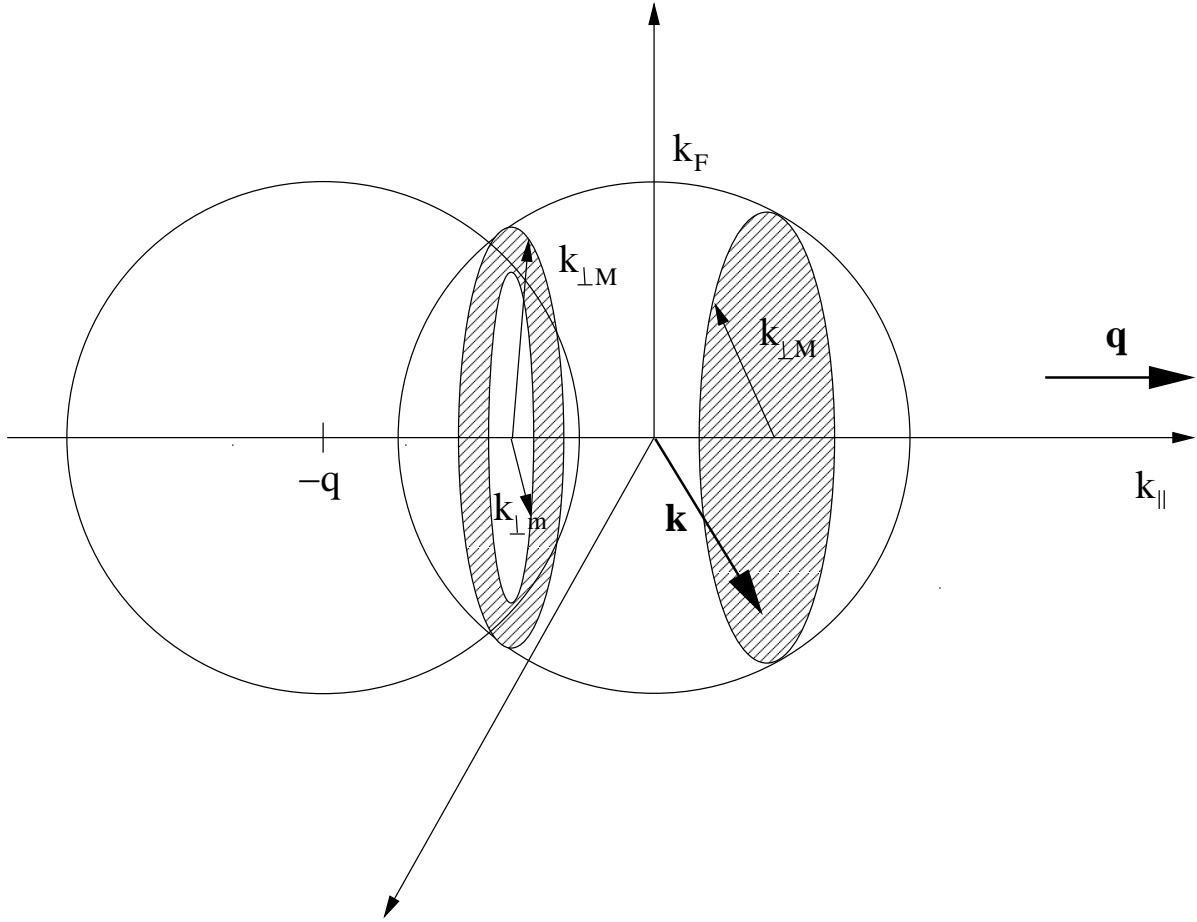}
\caption{The loci of the constant electron-hole excitation energies, containing
all wave vectors ${\bf k}$ with a same energy $E({\bf k},{\bf q})=E({\bf k}+
{\bf q})-E({\bf k})=(2k_{\parallel}+q)q/2m = const.$}
\label{loci}
\end{figure}

\begin{figure}
\vspace*{16.5cm}
\includegraphics{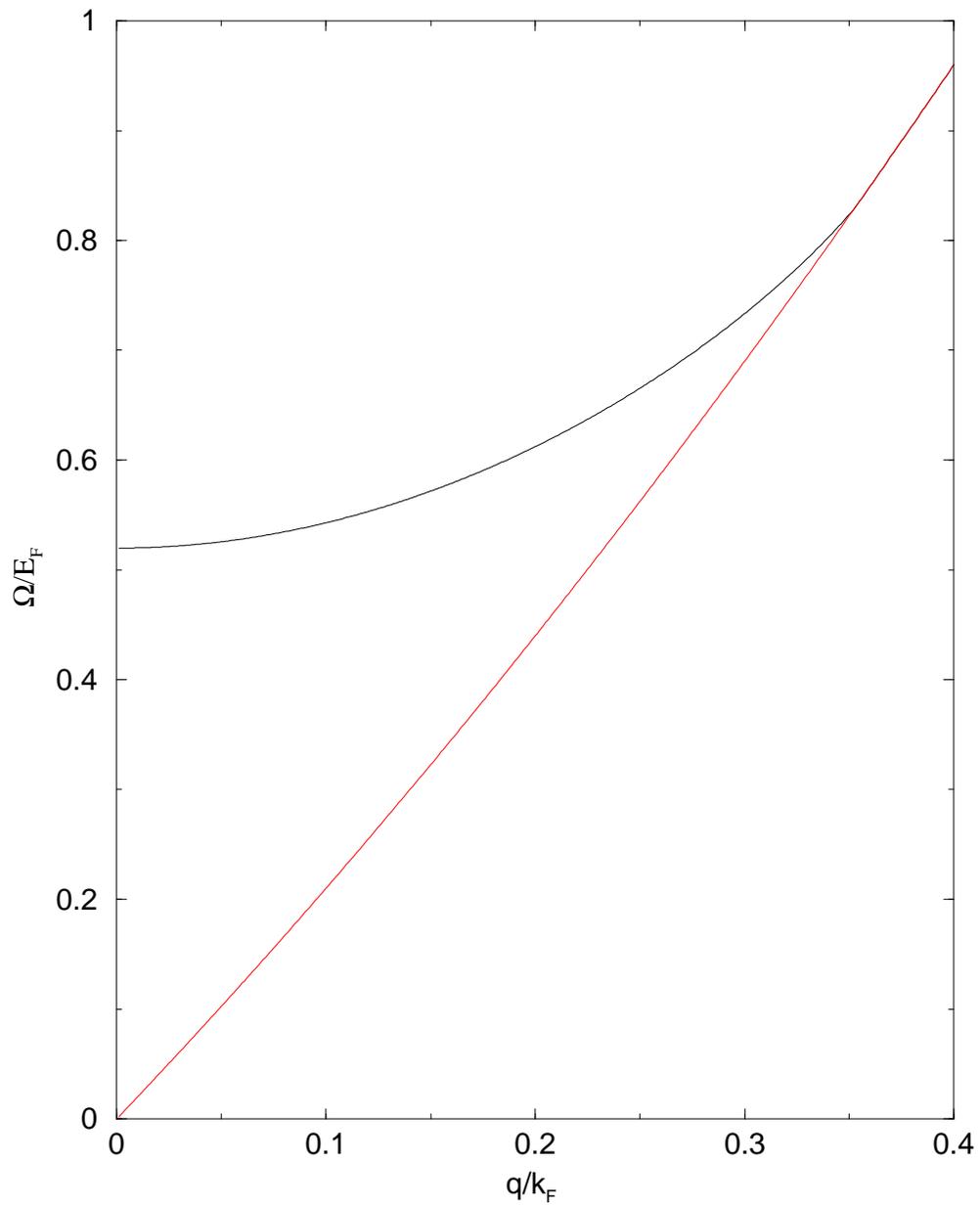}
\caption{The dispersion $\Omega (k_F,q)$ for $N=8.37 \cdot 10^9 $, and $L=10 a_0$.
 The asymptotic curve starting from origin is  $E(k_F,q)$. }
\label{plazmon}
\end{figure}

\begin{figure}
\vspace*{18cm}
\includegraphics{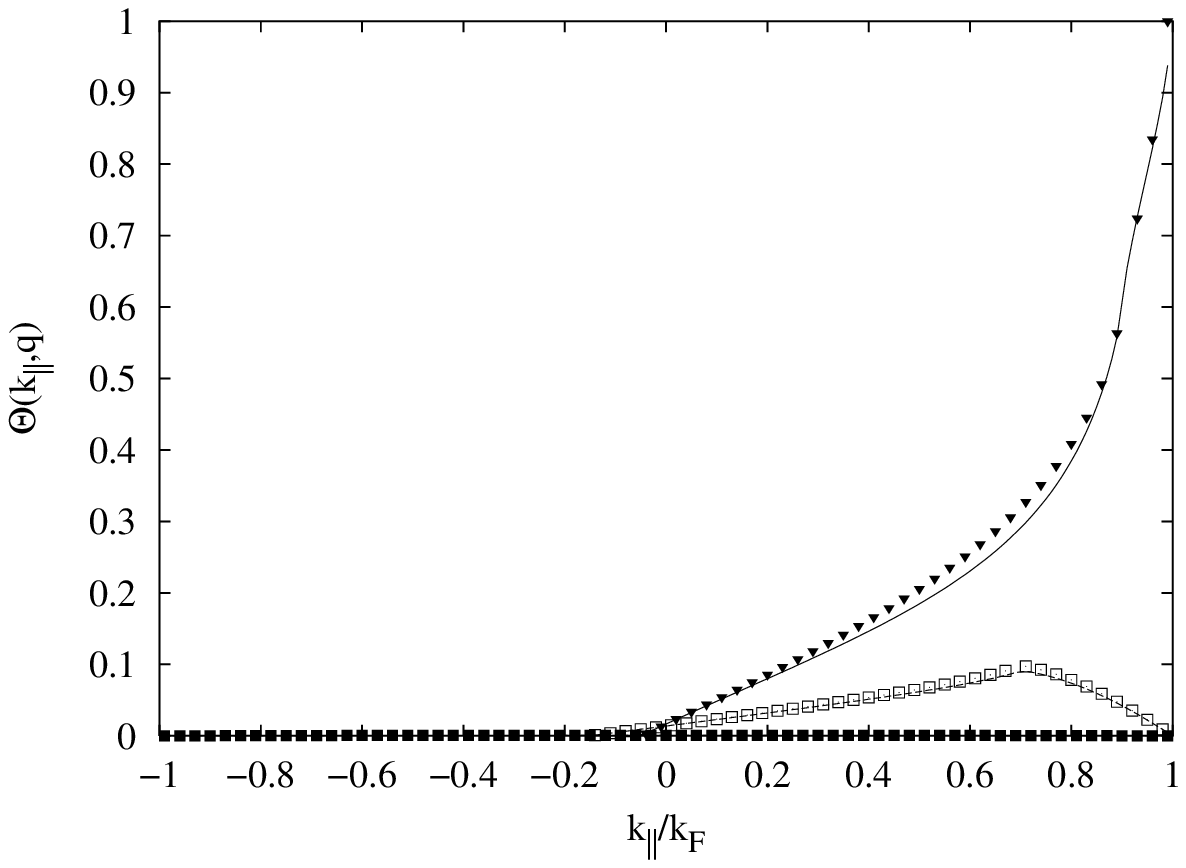}
\caption{The function $\Theta( k_{\parallel},{\bf q}) $ calculated by means of 
Eq. (\ref{difom}) (solid lines) and numerically from Eq. (\ref{dielectric1}) for 
$k_F=200\pi/L$, $L/a_0=10$ and for wave numbers $q = 0.1 k_F$ (triangles), 
$0.3k_F$ (hollow squares) and $2 k_F$ (filled squares).}
\label{test}
\end{figure}


\begin{references}
\bibitem{pinesel} See e. g. D. Pines {\em Elementary Excitations in Solids}, 
W. A. Benjamin, Inc., New York, Amsterdam, 1964.
\bibitem{doniach} S. Doniach and E. H. Sondheimer {\em Green's Functions for 
Solid State Physicist}, W. A. Benjamin, Inc., New York, Amsterdam, 1974.
\bibitem{zbb}P. \v{Z}upanovi\'{c}, A. Bjeli\v{s} and S. Bari\v{s}i\'{c}, 
Europhys. Lett. {\bf 45} (2), 188 (1999).
\bibitem{zb} P. \v{Z}upanovi\'{c} and  A. Bjeli\v{s}, to be published. 
\bibitem{zab} P. \v{Z}upanovi\'{c}, \v{Z}.  Agi\'{c} and A.  Bjeli\v{s}, to be 
published.
\bibitem{works} M. Gell-Mann and K.A. Brueckner, Phys. Rev. {\bf 106}, 
364 (1957); R. A. Ferrel, Phys. Rev. {\bf 107}, 450 (1957); K. Sawada, 
K. A. Brueckner, N. Fukuda and R. Brout, Phys. Rev. {\bf 108}, 507 (1957).
\bibitem{pinnoz} D. Pines and P. Nozi\`{e}res, {\em The Theory of Fermi 
Liquids}, W. A. Benjamin, Inc., New York, Amsterdam, 1966. 
\bibitem{fetter} A. L. Fetter and J. D. Walecka, {\em Quantum  Theory 
of Many-Particle Systems}, McGraw-Hill Book Company, New York, 1971.
\bibitem{mahan}G.D. Mahan {\em Many Particle Physics} Kluwer Academic/ Plenum 
 Publishers, New York, 2000.

\end{references}
\end{document}